\documentclass[aps,prb,floatfix,superscriptaddress,twocolumn,footinbib]{revtex4-2}

\usepackage[colorlinks=true, linkcolor=blue, filecolor=blue, urlcolor=blue, citecolor=blue]{hyperref}
\usepackage{amsfonts}
\usepackage{bbm}
\usepackage{graphicx}% Include figure files
\usepackage{dcolumn}% Align table columns on decimal point
\usepackage{bm}% bold math
\usepackage{soul}
\usepackage[version=3]{mhchem}
\usepackage{relsize}
\usepackage{mathrsfs}
\usepackage{physics}

\begin{document}

\title{Spin transport in a normal metal-altermagnetic superconducting nanowire junction}

\author{Xing-Jian Yi}
\affiliation{International Center for Quantum Materials, School of Physics, Peking University, Beijing 100871, China}
\affiliation{Hefei National Laboratory, Hefei 230088, China}

\author{Yi-Xin Dai}
\affiliation{International Center for Quantum Materials, School of Physics, Peking University, Beijing 100871, China}
\affiliation{Beijing Academy of Quantum Information Sciences, West Bld.\#3, No.10 Xibeiwang East Rd., Haidian District, Beijing 100193, China}

\author{Yue Mao}
\email[]{maoyue@pku.edu.cn}
\affiliation{International Center for Quantum Materials, School of Physics, Peking University, Beijing 100871, China}

\author{Qing-Feng Sun}
\email[]{sunqf@pku.edu.cn}
\affiliation{International Center for Quantum Materials, School of Physics, Peking University, Beijing 100871, China}
\affiliation{Hefei National Laboratory, Hefei 230088, China}
\affiliation{Beijing Academy of Quantum Information Sciences, West Bld.\#3, No.10 Xibeiwang East Rd., Haidian District, Beijing 100193, China}

\begin{abstract}
Spin-triplet superconductors are considered as a promising platform for dissipationless spin transport, where spin currents are carried by spin-triplet Cooper pairs. In this paper, we propose that the spin-triplet superconductivity and spin supercurrent can be engineered in an altermagnetic superconducting nanowire, where a one-dimensional (1D) nanowire is placed on the surface of an $s$-wave superconductor and in proximity to the altermagnet. Using the nonequilibrium Green's function method, we demonstrate a nonzero equal-spin Andreev reflection coefficient at the normal metal-altermagnetic superconducting nanowire interface, thereby verifying the injection of spin-triplet Cooper pairs. %in such a system.
Furthermore, we systematically investigate the spin transport properties in this hybrid system under a spin bias. Our results demonstrate that these properties %spin conductance and spin injection efficiency
can be effectively tuned by the chemical potential and spin bias orientation. Our proposal provides a pathway toward realizing dissipationless spin transport.

\end{abstract}
\maketitle

\section{\label{sec1}Introduction}

Spintronics has evolved into a persistently active research field in condensed matter physics over the past two decades, owing to its promising applications in data storage and logical operations controlled by spin currents \cite{sinova_New_2012, zutic_Spintronics_2004, wolf_Spintronics_2001}.
Compared with the manipulation of charge degree of freedom, the spin degree of freedom exhibits faster switching speed and significantly lower energy consumption.
In spintronics, the generation and control of spin currents constitute a fundamental research focus. For example, a pure spin current can be generated through two primary mechanisms: the bulk spin Hall effect in heavy metals and the interfacial Rashba-Edelstein effect at symmetry-broken interfaces \cite{cao_Pure_2025, liu_SpinTorque_2012, sinova_Spin_2015}.
For practical applications, spin-transfer torque and spin-orbit torque have emerged as two dominant approaches for efficient spin state control, enabling their widespread applications in magnetic tunnel junction and magnetoresistive random-access memory \cite{baibich_Giant_1988, cubukcu_Spinorbit_2014, fukami_Magnetization_2016, ikeda_Perpendicularanisotropy_2010, manchon_Currentinduced_2019, mellnik_Spintransfer_2014, ralph_Spin_2008, yuasa_Giant_2004, zhang_Mechanisms_2002}.
However, spin relaxation inevitably degrades spin transport efficiency in practical device operations.
This relaxation process originates from multiple scattering mechanisms, including spin-orbit coupling (SOC), elastic spin flips by magnetic impurities, the D'yakonov-Perel' interaction and phonon Raman scattering, all of which could reduce the lifetime of spin currents and limit the performance of spintronic devices \cite{jarmola_Temperature_2012, bobkova_Longrange_2015, bobkova_Injection_2016, heikkila_Thermal_2019, krishtop_Nonequilibrium_2015, silaev_LongRange_2015, silaev_Spin_2015, lundeberg_DefectMediated_2013, tackeuchi_Electron_1997, zihlmann_Large_2018}.

In superconducting systems, electrons condense into Cooper pairs and form a coherent superfluid state, enabling the dissipationless transport of charge current. However, conventional superconductors (SCs) host spin-singlet Cooper pairs with no spin polarization, preventing spin transport mediated by Cooper pairs.
The discovery of spin superconductors and spin-triplet superconductors provides a potential pathway to overcome this restriction.
In spin superconductors, the condensation of spin-triplet electron-hole pairs can form a superfluid state that enables dissipationless spin transport, whereas the whole system remains a charge insulator \cite{sun_Spin_2011, sun_Spinpolarized_2013, nakata_Josephson_2014, bozhko_Supercurrent_2016}.
On the other hand, the discovery of spin-triplet superconductors enables the investigation of Cooper-pair-mediated spin transport in superconducting systems \cite{mackenzie_Superconductivity_2003}.
Many platforms have been proposed for realizing spin-triplet superconductor, such as superconductor-ferromagnet structures \cite{bergeret_Odd_2005}, 
two-dimensional Ising superconductors \cite{zhou_Ising_2016, lv_Magnetoanisotropic_2018, mockli_Magneticfield_2019}, as well as one-dimensional (1D)
superconducting nanowire with SOC and external magnetic field \cite{lutchyn_Majorana_2010, oreg_Helical_2010a, mao_Spin_2022, mao_Universal_2024}.
When an electron is incident on the spin-triplet superconductor, it can be reflected as a hole within the same spin sub-band via equal-spin Andreev reflection.
This process corresponds to the injection of equal-spin Cooper pairs into the spin-triplet superconductor, thereby establishing a dissipationless spin superfluid \cite{visani_Equalspin_2012}.

Altermagnet has recently emerged as a novel magnetic phase distinct from conventional ferromagnet and antiferromagnet, characterized by spin-splitting band structures in momentum space and vanishing net magnetization in real space \cite{smejkal_Conventional_2022a, smejkal_Emerging_2022a}.
This unique phase has been experimentally observed in various material systems \cite{fedchenko_Observation_2024a,krempasky_Altermagnetic_2024a,reimers_Direct_2024a}.
Altermagnet phase also exhibits intriguing transport phenomena including the spin splitter effect \cite{gonzalez-hernandez_Efficient_2021a}, crystal Hall effect \cite{smejkal_Crystal_2020} and  giant and tunneling magnetoresistance effects \cite{smejkal_Giant_2022,sun_Tunneling_2025}. Besides, its integration with superconductivity, thermoelectrics, topological states and multiferroics has demonstrated outstanding research potential \cite{cheng_Fieldfree_2024a,ghorashi_Altermagnetic_2024a,li_Majorana_2023a,wan_Altermagnetisminduced_2025a,yi_Spin_2025a,duan_Antiferroelectric_2025,smejkal_Emerging_2022a, wan_Interplay_2025,addr1}.
As a magnetic phase, altermagnetism breaks spin-rotational symmetry.
Therefore, when combined with superconductivity, it can also exhibit spin-triplet superconductivity \cite{zhu_Topological_2023,wei_gapless_2024,decarvalho_unconventional_2024,maeda_Classification_2025,Chakraborty_Constraints_2025}, allowing for the injection of spin supercurrents.

\begin{figure}[!htb]
    \centerline{\includegraphics[width=\columnwidth]{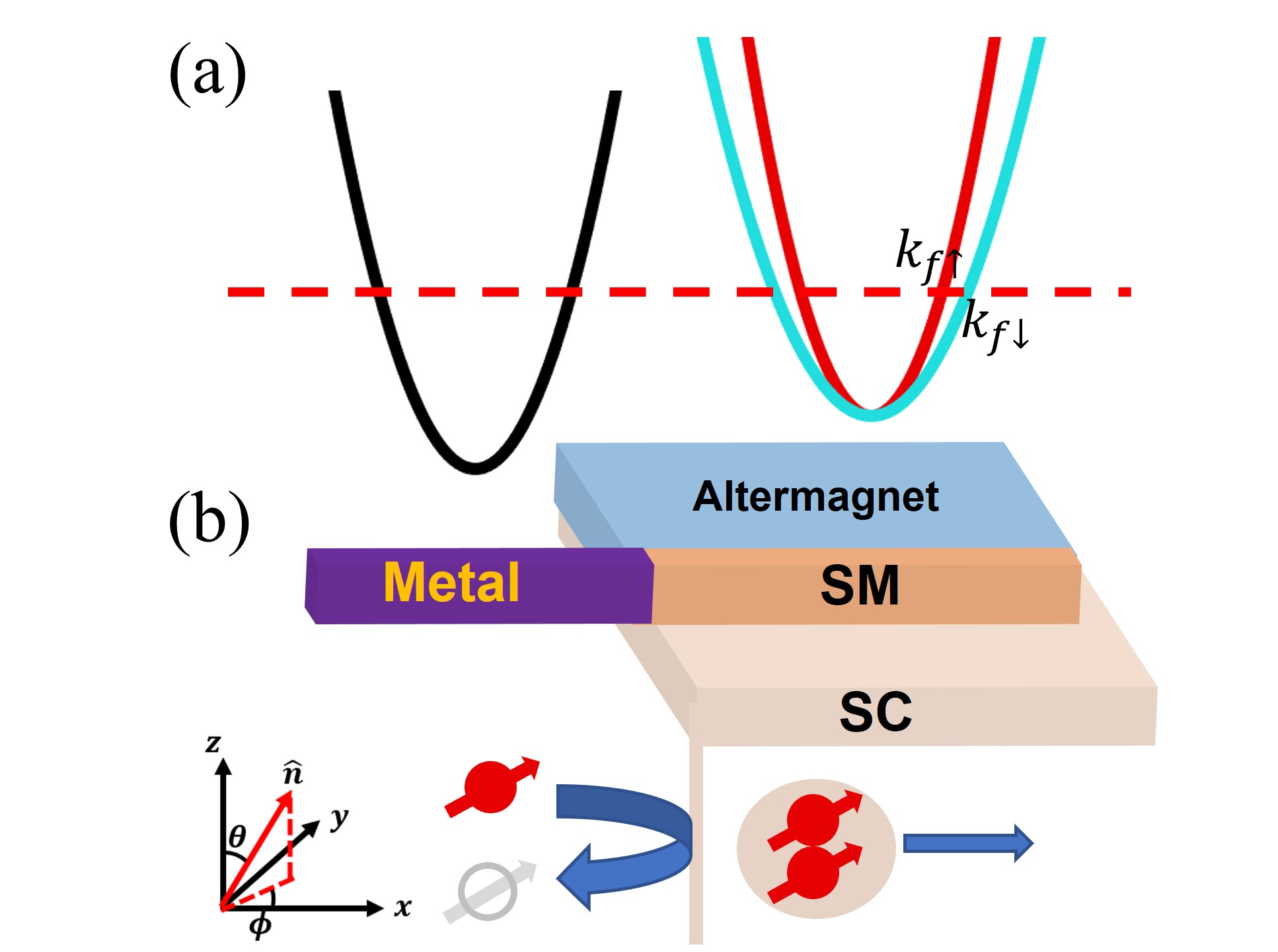}}
    \caption{Schematic for normal metal-altermagnetic superconducting nanowire junction. (a) The left is the schematic energy band of normal metal. The right is the schematic energy band of altermagnet accompanied by spin splitting. The red dashed line is Fermi level. $k_{f\uparrow}$ and $k_{f\downarrow}$ are Fermi wave vectors for spin-up (red) and spin-down (cyan) bands, respectively.
    (b) The upper part is the schematic of the device, where a metal connects with a nanowire on the surface of an $s$-wave SC and in proximity to altermagnet. The lower part is the spin transport process. %Under an $\hat{\bm{n}}$ direction spin bias, a spin supercurrent is injected into the altermagnetic superconducting nanowire by equal-spin Andreev reflection.
    When a spin-$\hat{\bm{n}}$ electron is incident, it is reflected as an equal-spin hole, injecting a spin-triplet Cooper pair into the superconducting nanowire.
    }\label{Fig1}
\end{figure}

In this paper, we investigate the spin transport in a normal metal-1D altermagnetic superconducting nanowire junction. The altermagnetic superconducting nanowire is realized by a nanowire placed on the surface of an $s$-wave SC and in proximity to the altermagnet, as illustrated in Fig. \ref{Fig1}.
Using the pairing correlation function, we analyze the existence of the equal-spin pairing component theoretically in two cases, whether SOC is present or not. Remarkably, even in the absence of SOC, a substantial equal-spin pairing component persists with the spin orientation perpendicular to the Néel vector of the altermagnet.
Using nonequilibrium Green's function method, we confirm the existence of the equal-spin pairing through the nonzero equal-spin Andreev reflection coefficient.
We next investigate the spin transport properties under a spin bias, which can be experimentally applied and regulated by the spin Hall effect and ferromagnetic resonance spin pumping \cite{hirsch_Spin_1999, tserkovnyak_Spin_2002}. We show that spin conductance and spin injection efficiency from the metal into the superconductor can be well tuned by chemical potential and spin bias orientation.
In the presence of SOC, the altermagnetic superconducting nanowire can be tuned into a topological superconducting state, with the spin conductance reaching a quantized value.

The rest of the paper is organized as follows.
In Sec. \ref{sec2},
the tight-binding lattice Hamiltonian for the normal metal-altermagnetic superconducting nanowire junction is given. The formulas for calculating the reflection coefficient, spin current, spin conductance and spin injection efficiency are then presented and the pairing correlation function is analyzed.
Sec. \ref{sec3} gives the results of equal-spin Andreev reflection coefficient, spin conductance and spin injection efficiency in the absence of SOC.
Sec. \ref{sec4} gives corresponding results in the presence of SOC.
Finally, a discussion and summary is given in Sec. \ref{sec5}.

\section{\label{sec2}Mechanism and methods}

As illustrated in Fig. \ref{Fig1}, a semi-infinite metal lead is contacted to the 1D altermagnetic superconducting nanowire. The fabrication and probing of such nanowire device are mature in experiments, e.g. see Ref. \cite{mourik_Signatures_2012,wang_singlet_2022}. In the low-energy (expanded near the $\Gamma$ point), the nanowire can be described by the Bogoliubov-de Gennes (BdG) Hamiltonian $H^{\mathrm{BdG}}$ in the Nambu basis $\Psi=\{\psi_{k_x,\uparrow}, \psi_{k_x,\downarrow},\psi_{-k_x,\uparrow}^\dagger ,\psi_{-k_x,\downarrow}^\dagger \}^T$:
\begin{align}
H^{\mathrm{BdG}}=(\frac{\hbar^2 k_x^2}{2m^*}-\mu+T_j k_x^2 \sigma_z+\alpha_R k_x \sigma_y)\tau_z-\Delta \sigma_y \tau_y,
\label{E1}
\end{align}
where $k_x$ is the wave vector of $x$ direction, $m^*$ is the effective electron mass, $\mu$ is the chemical potential, $\sigma_{x,y,z}$ and $\tau_{x,y,z}$ are Pauli matrices acting on spin space and particle-hole space, respectively. $T_j$ is the proximitized altermagnet strength with its Néel vector along $z$ direction, $\alpha_R$ is the SOC strength and $\Delta$ is the proximity-induced $s$-wave pairing potential. 
$\Delta$ and $T_j$ are induced by the proximity from superconductor and altermagnet, respectively \cite{stanescu_proximity_2010,volkov_Proximity_1995,zhu_Altermagnetic_2025}.
They are usually caused by weak coupling, and the induced effects are weaker than those in bulk systems, but still significant.
%Meanwhile, the weak coupling prevents electrons from transferring between nanowire, superconductor, and altermagnet.
%Meanwhile, the weak coupling effect makes the electron transfer between the nanowire, superconductor, and altermagnet so weak that it can be neglected \cite{???}. 
Additionally, under a spin bias, the current flowing into 
the altermagnetic superconducting nanowire is a spin supercurrent, carried by spin-triplet Cooper pairs. 
These spin-triplet Cooper pairs are completely prohibited from transferring to the non-superconducting altermagnet and $s$-wave superconductors. 
Therefore, the altermagnet and $s$-wave superconductors [see Fig. \ref{Fig1}(b)]
can be neglected in our Hamiltonian in Eq. (\ref{E1}).
To facilitate numerical calculations, we use tight-binding lattice Hamiltonian \cite{ghorashi_Altermagnetic_2024a} and couple the metal lead:
\begin{align}
H=H_{nw}+H_{lead}+H_{couple},
\label{E2}
\end{align}
where
\begin{align}
H_{nw}=&\sum_{i>0,s} (2t-\mu)c_{i,s}^\dagger c_{i,s}-\sum_{i>0,s} t(c_{i,s}^\dagger c_{i+1,s}+H.c.) \nonumber \\
&+\sum_{i>0,s,s'} 2t_Jc_{i,s}^\dagger(\sigma_z)_{s,s'} c_{i,s'}  \nonumber \\
&-\sum_{i>0,s,s'} t_J (c_{i,s}^\dagger(\sigma_z)_{s,s'} c_{i+1,s'}+H.c.) \nonumber \\
&-\sum_{i>0,s,s'}\frac{\alpha}{2}(c_{i,s}^\dagger(i\sigma_y)_{s,s'} c_{i+1,s'}+H.c.) \nonumber \\
&+\sum_{i>0} (\Delta c_{i,\uparrow}^\dagger c_{i,\downarrow}^\dagger+H.c.)
,
\label{E3}
\end{align}
and
\begin{align}
&H_{lead}=\sum_{i\leq 0,s} t a_{i,s}^\dagger a_{i,s}-\sum_{i\leq 0,s} t(a_{i,s}^\dagger a_{i-1,s}+H.c.), \nonumber \\
&H_{couple}=-\sum_st_c(a_{0,s}^\dagger c_{1,s}+H.c.)
.
\label{E4}
\end{align}
$H_{nw}$ is the altermagnetic superconducting nanowire Hamiltonian, which is used for all subsequent calculations. Based on Eq. (\ref{E3}), we can derive its low-energy effective Hamiltonian, given in Eq. (\ref{E1}). The corresponding parameters are related as $t=\hbar^2/{2m^*a^2}$, $t_J=T_j/a^2$ and $\alpha=\alpha_R/a$ ($a$ is lattice constant). $c_{i,s}^\dagger$ $(c_{i,s})$ is the creation (annihilation) operator of electron with spin $s$ on lattice site $i>0$ of the superconducting nanowire.
$H_{lead}$ is the metal lead Hamiltonian, obtained by just setting $t_J=0$, $\alpha=0$, $\Delta=0$, and $\mu=t$ in Eq. (\ref{E3}).
$a_{i,s}^\dagger$ $(a_{i,s})$ is the creation (annihilation) operator of electron with spin $s$ on lattice site $i\leq0$ of metal lead.
$H_{couple}$ is the point coupling between superconducting nanowire and metal lead with $t_c$ the hopping strength.

Under a spin bias of $\hat{\bm{n}}$ direction in normal metal lead, the spin current in normal lead can be obtained by \cite{wang_Spinbattery_2004}
\begin{align}
J_s=\frac{\hbar}{2}(J_+-J_-),
\label{E5}
\end{align}
with the corresponding basis $|\pm\rangle = | \hat{\bm{n}} \cdot \vec{\mathbf{S}} = \pm \hbar/2 \rangle$. Here $\hat{\bm{n}}=(\sin \theta \cos\phi,\sin\theta\sin\phi,\cos\theta)$, where $\theta$ and $\phi$ are polar angle and azimuthal angle [see Fig. \ref{Fig1}(b)], respectively.
We obtain each specific spin current $J_{\sigma}(\sigma = \pm)$ by using Landauer-B$\ddot{u}$ttiker formula and nonequilibrium Green's function method \cite{lv_Magnetoanisotropic_2018,dai_Spin_2022a,meir_Landauer_1992},
\begin{align}
J_{\sigma}=
\frac{1}{h} \int d E &[(f_{L,e\sigma}-f_{L,e\bar{\sigma}}) T_{{\bar e}e,\sigma} +(f_{L,e\sigma}-f_{L,h {\sigma}})T_{h e ,\sigma} \nonumber \\
&+ (f_{L,e\sigma}-f_R)  T_{trans ,\sigma}],
\label{E6}
\end{align}
where $h$ is the Planck constant and $E$ is the energy.
$T_{{\bar e}e,\sigma}=\mathrm{Tr}[\bm{\Gamma}_{ee\sigma\sigma}^L \mathbf{G}_{ee\sigma \bar{\sigma}}^r  \bm{\Gamma}^L_{ee\bar{\sigma}\bar{\sigma}} \mathbf{G}_{ee\bar{\sigma} \sigma}^a]$ is the spin-flip reflection coefficient where an electron is reflected to an opposite-spin electron \cite{lv_Spinflip_2017a}, $T_{h e ,\sigma}=\mathrm{Tr}[\bm{\Gamma}_{ee\sigma\sigma}^L \mathbf{G}_{eh\sigma \sigma}^r  \bm{\Gamma}_{hh\sigma\sigma}^L \mathbf{G}_{he\sigma \sigma}^a]$ is the equal-spin Andreev reflection coefficient where an electron is reflected to an equal-spin hole \cite{he_Selective_2014a}, and $T_{trans ,\sigma}=\mathrm{Tr} [ \bm{\Gamma}_{ee\sigma\sigma}^L (\mathbf{G}^r  \bm{\Gamma}^R \mathbf{G}^a)_{ee\sigma \sigma}]$ is the
transmission coefficient corresponding to the quasiparticle tunneling.
{The normal Andreev reflection coefficient $T_{{\bar h}e,\sigma}=\mathrm{Tr}[\bm{\Gamma}_{ee\sigma\sigma}^L \mathbf{G}_{eh\sigma \bar{\sigma}}^r  \bm{\Gamma}_{hh\bar{\sigma}\bar{\sigma}}^L \mathbf{G}_{he\bar{\sigma} \sigma}^a]$ and normal reflection coefficient $T_{ee,\sigma}=1-T_{{\bar e}e,\sigma}-T_{h e ,\sigma}-T_{{\bar h}e,\sigma}-T_{trans ,\sigma}$ do not contribute to the spin current.}
$\mathbf{G}^{r(a)}$ is retarded (advanced) Green's functions of center region and $\bm{\Gamma}^{L/R}$ are linewidth matrices for left metal lead and right superconducting nanowire, both are expressed in BdG basis. Here we have chosen the outmost site of the normal-metal lead to be the center region. Note that the selection of the center region has no impact on the calculation results.
In the normal lead, $f_{L,e+}=f_{L,h-}=f(E-eV)$ and $f_{L,e-}=f_{L,h+}=f(E+eV)$ with $V$ the spin bias which drives opposite-spin electrons oppositely \cite{wang_Spinbattery_2004, mao_Charge_2021a}.
In the superconducting nanowire, $f_{R}=f(E)$ and $f(E)=1/\{exp[(E-E_F)/k_B \mathcal{T}]+1  \}$ is the Fermi distribution function.
$\mathcal{T}$ is the temperature and $E_F$ is the Fermi level, both of which are set to zero.
By substituting Eq. (\ref{E6}) in Eq. (\ref{E5}), the differential spin conductance $G_{s,tot}=dJ_s/dV$ in the low-temperature limit is given by
\begin{align}
    G_{s,tot}=&\frac{e}{4\pi}\sum_{\sigma=\pm}[T_{{\bar e}e,\sigma}(eV)+T_{{\bar e}e,\sigma}(-eV)\nonumber \\
    &+T_{h e ,\sigma}(eV)+T_{h e ,\sigma}(-eV)]\nonumber \\
    &+\frac{e}{4\pi}[T_{trans ,+}(eV)+T_{trans ,-}(-eV)] .
    \label{E7}
\end{align}
In calculation, we set $\frac{e}{4\pi}$ as spin conductance unit.
The spin current is defined in normal lead so that three processes contribute to the total spin conductance: spin-flip reflection, equal-spin Andreev reflection and quasiparticle tunneling, but spin current contributed by spin-flip reflection process dissipates at the interface and does not inject into the altermagnetic superconducting nanowire \cite{dai_Spin_2022a}. Thus we define a spin injection efficiency as
\begin{align}
    \eta&=1-G_{s,dis}/G_{s,tot}, \nonumber \\
     G_{s,dis}&=\frac{e}{4\pi}\sum_{\sigma=\pm}[T_{{\bar e}e,\sigma}(eV)+T_{{\bar e}e,\sigma}(-eV)]
    ,
    \label{E8}
\end{align}
where $G_{s,dis}$ is the dissipated part contributed by spin-flip reflection process.

To investigate the spin-triplet pairing symmetry at energy $E$ in the superconducting nanowire, we use pairing correlation functions \cite{balian_Superconductivity_1963,tang_Magnetic_2021a,zhou_Ising_2016}, which is obtained for an infinite wire described by the effective Hamiltonian in Eq. (\ref{E1})
\begin{align}
    F_{\sigma\sigma'}(k_x,E)=-i\int_0^{+\infty} dt e^{i(E+i0^+)t} \langle \{\psi_{k_x,\sigma}(t), \psi_{-k_x,\sigma'}(0)\} \rangle
    .
    \label{E9}
\end{align}
$\psi_{k_x,\sigma}$ is the annihilation operator for spin $\sigma$.
We obtain $\mathbf{F}$ in $2\times2$ matrix form.
This matrix can be decomposed as $\mathbf{F}=(d_0 \sigma_0+{\bm{d}}\cdot{\bm{\sigma}}) i\sigma_y$, where $d_0$ represents the spin-singlet pairing and ${\bm{d}}$ vector represents spin-triplet pairing. We find
\begin{align}
{\bm{d}}=\frac{2\Delta}{M}(iAJ,AT,J E) ,
\label{E10}
\end{align}
where $M=(A^2+E^2+J^2-\Delta^2-T^2)^2+4(A^2 \Delta^2-A^2E^2-J^2E^2)$. $A=\alpha_R  k_x$, $J=T_j k_x^2$ and $T=\frac{\hbar^2 k_x^2}{2m^*} -\mu$.
$A$, $J$ and $T$ respectively correspond to orbit component of SOC term, altermagnet splitting term and kinetic energy term in Eq. (\ref{E1}).
This ${\bm{d}}$ vector indicates that the altermagnet and SOC induce effective spin-triplet pairing correlations, which are analyzed in detail in Secs. \ref{sec3}, \ref{sec4}.

In the numerical calculations, we set effective mass $m^*=0.02m_e$, induced superconducting pairing strength $\Delta=0.25$ meV, $\alpha_R=0.28$ eV $\mathrm{\mathring{A}}$ \cite{mourik_Signatures_2012,zhuang_Anomalous_2022a}. The lattice constant $a$ is set as $10$ nm. We set $\Delta$ as energy unit. Consequently, the hopping energy $t=\hbar^2/2m^* a^2 \approx 20$ meV, corresponding to $80\Delta$. $\alpha=\alpha_R/a=2.8$ meV, corresponding to $11.2\Delta$. The hopping between metal lead and superconducting nanowire $t_c$ is set as $0.5t$, corresponding to $40\Delta$. Unless noted in figure captions, these parameter values are used throughout the paper.

\section{\label{sec3}Spin transport in the absence of SOC}

\begin{figure}[!htb]
\centerline{\includegraphics[width=\columnwidth]{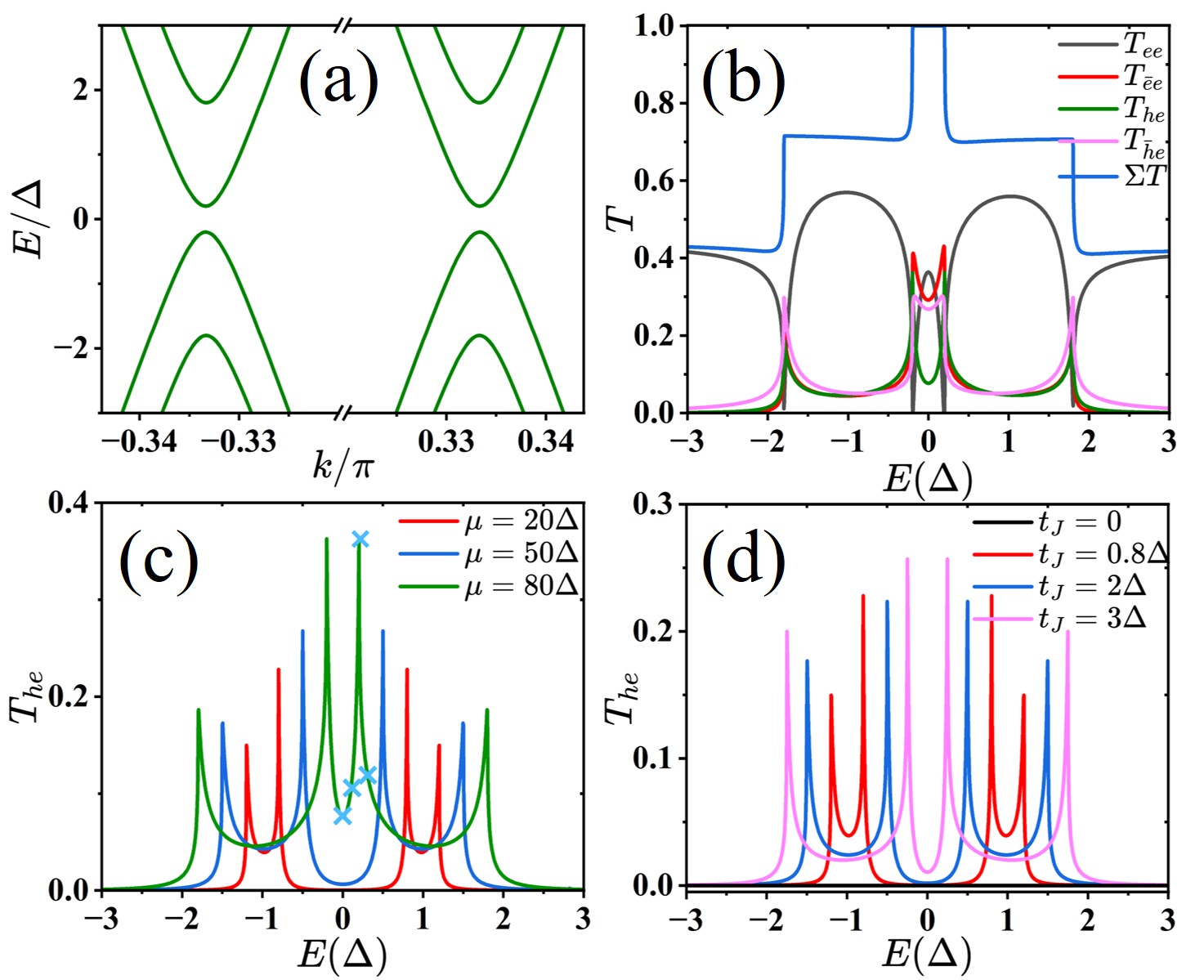}}
\caption{(a) The energy band of 1D superconducting nanowire. (b) Four reflection coefficients and total reflection coefficient versus incident electron energy $E$. (c, d) Equal-spin Andreev
reflection coefficient $T_{he}$ versus incident electron energy $E$ for (c) different chemical potentials $\mu$ and (d) different altermagnet strengths $t_J$. Parameters:  $\alpha=0$, $\theta=\pi/2$, and $\phi=0$ in (a-d), $t_J=0.8\Delta$ in (a-c), $\mu=80\Delta$ in (a-b), $\mu=20\Delta$ in (d).
}\label{Fig2}
\end{figure}

\begin{figure}[!htb]
    \centerline{\includegraphics[width=\columnwidth]{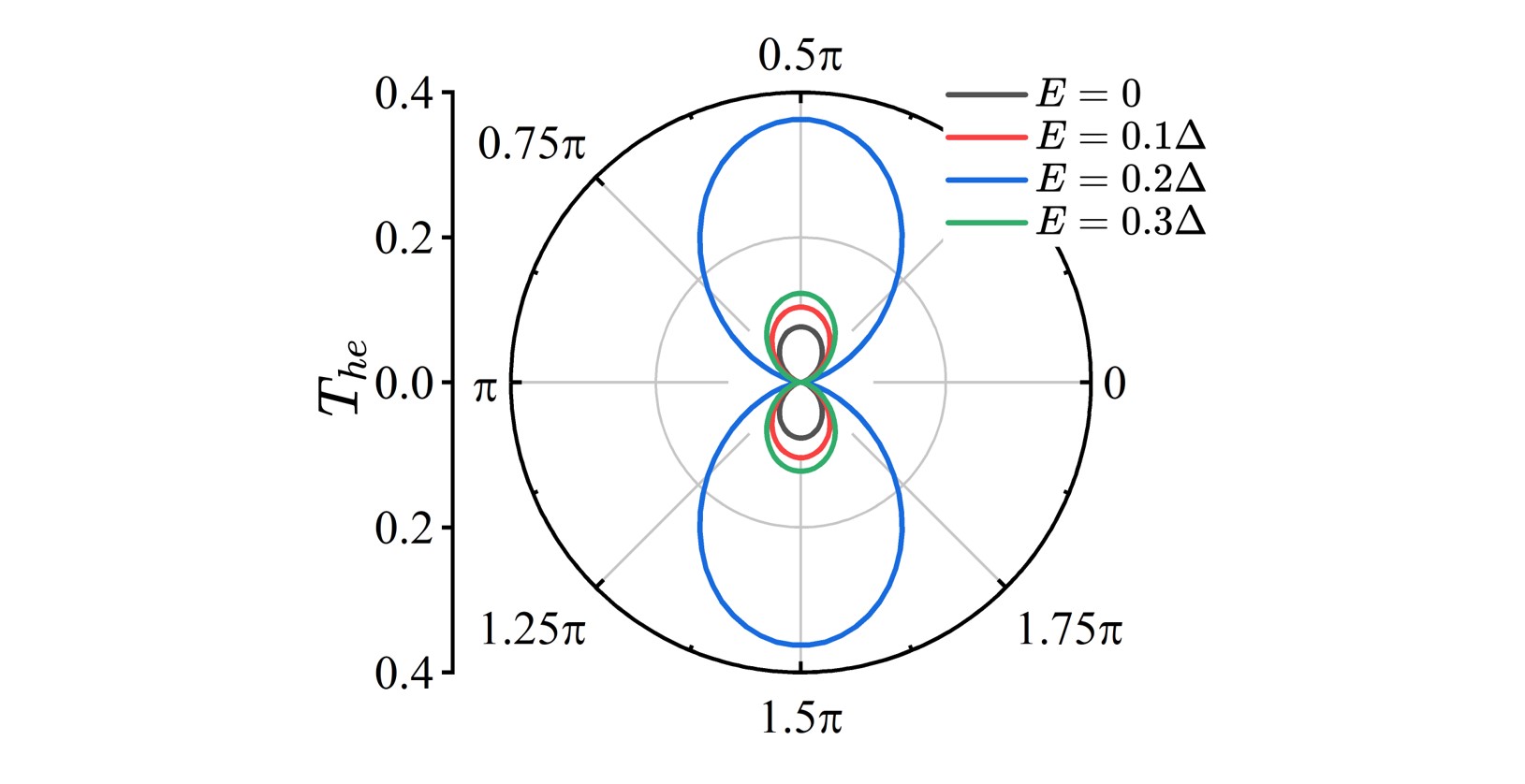}}
    \caption{Equal-spin Andreev reflection coefficient $T_{he}$ versus polar angle $\theta$ for different incident electron energies $E$. The four data points at $\theta=\pi/2$ on four curves correspond to the four blue cross points in Fig. \ref{Fig2}(c). Parameters: $\alpha=0$, $t_J=0.8\Delta$, $\mu=80\Delta$, $\phi=0$.
    }\label{Fig3}
\end{figure}

In this section, we investigate the spin transport characteristic without SOC by setting $\alpha=0$ in 1D nanowire. Through Eq. (\ref{E10}), we obtain ${\bm{d}}=(0,0,d_z)$ because the SOC term $A=0$.
The ${\bm{d}}$ vector has only $z$ component, indicating the spin-triplet pairing in superconducting nanowire takes the form $|S=1, S_z=0\rangle=\frac{1}{\sqrt{2}}(\ket{\uparrow\downarrow}+\ket{\downarrow\uparrow})_z$ \cite{tang_Magnetic_2021a}.
In the spin basis along $z$ direction,
the Cooper pairs are formed by electrons with opposite spin.
Thus, they do not contribute to spin transport when spin bias is along $z$ direction.
However, by choosing an arbitrary orientation of the spin quantization axis $\hat{\bm{n}}=(\sin \theta \cos\phi,\sin\theta\sin\phi,\cos\theta)$, along with $\ket{\uparrow}_z=\cos \frac{\theta}{2} \ket{\uparrow}_{\hat{\bm{n}}}-\sin \frac{\theta}{2} e^{-i\phi}\ket{\downarrow}_{\hat{\bm{n}}}$ and $\ket{\downarrow}_z=\sin \frac{\theta}{2} e^{i\phi}\ket{\uparrow}_{\hat{\bm{n}}}+\cos \frac{\theta}{2} \ket{\downarrow}_{\hat{\bm{n}}}$,
the triplet pairing can be expressed by
\begin{align}
\big(\ket{\uparrow\downarrow}+\ket{\downarrow\uparrow}\big)_z&=\sin \theta \big(e^{i\phi}\ket{\uparrow\uparrow}_{\hat{\bm{n}}} -e^{-i\phi}\ket{\downarrow\downarrow}_{\hat{\bm{n}}}\big)   \nonumber  \\
&+\cos\theta\big(\ket{\uparrow\downarrow}+\ket{\downarrow\uparrow}\big)_{\hat{\bm{n}}} ,
\label{E11}
\end{align}
where equal-spin pairing Cooper pairs emerge as long as $\theta \neq 0$.
Besides, equal-spin pairing component reaches its maximum at $\theta=\pi/2$, where the spin orientation is perpendicular to the direction of Néel vector of the altermagnet ($z$ axis).

\begin{figure*}[!htb]
\centerline{\includegraphics[width=1.8\columnwidth]{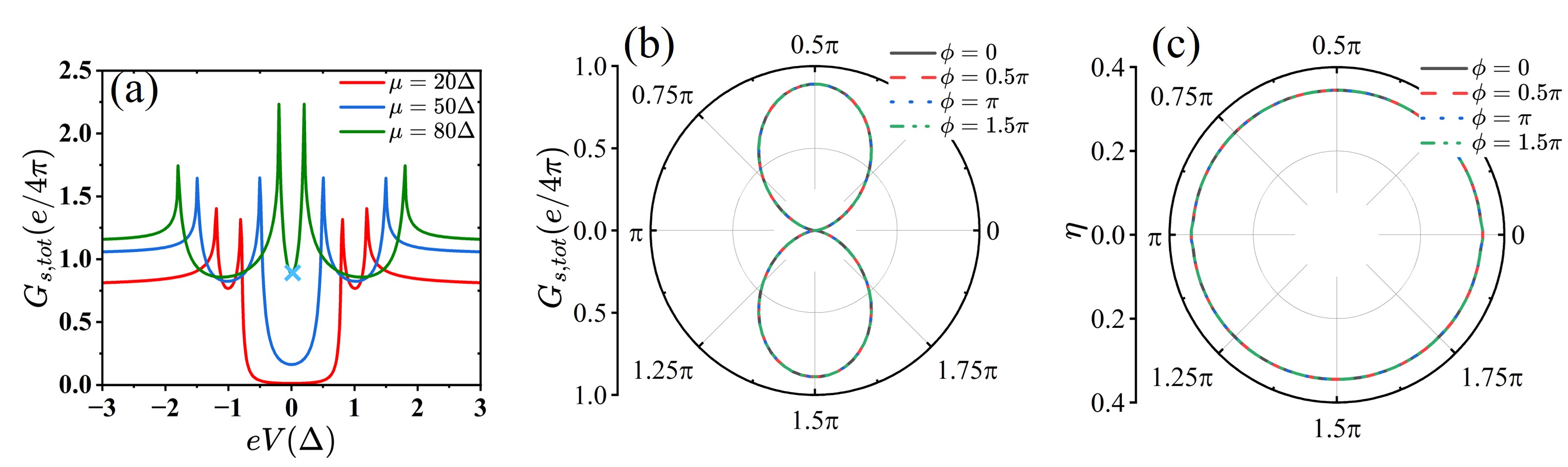}}
\caption{
    (a) Spin conductance $G_{s,tot}$ versus spin bias $V$ for different chemical potentials $\mu$. (b) Spin conductance $G_{s,tot}$ versus polar angle $\theta$ for different azimuthal angles $\phi$. Point at $\theta=\pi/2$ on $\phi=0$ curve corresponds to blue cross point in (a). (c) Spin injection efficiency $\eta$ versus polar angle $\theta$ for different azimuthal angles $\phi$. Parameters: $\alpha=0$ and $t_J=0.8\Delta$ in (a-c), $\mu=80\Delta$ and $V=0$ in (b-c), $\theta=\pi/2$ and $\phi=0$ in (a).
}\label{Fig4}
\end{figure*}

To verify and quantitatively characterize this effect, we calculate equal-spin Andreev reflection coefficient in Figs. \ref{Fig2} and \ref{Fig3}.
Based on our analysis, the equal-spin Andreev reflection coefficient should exhibit a direction dependence, specifically determined by the spin orientation of incident electron.
In Fig. \ref{Fig2}, we select the spin of incident electron along $x$ direction as a representative case.
As illustrated in Fig. \ref{Fig2}(a), in the presence of altermagnet, the BdG energy band demonstrates spin splitting as four bands. Fig. \ref{Fig2}(b) shows four reflection coefficients and total reflection coefficient as functions of incident electron energy,
{where the total reflection coefficient is obtained by $\sum T=T_{ee}+T_{{\bar e}e}+T_{h e }+T_{{\bar h}e}$.}
The nonzero equal-spin Andreev reflection coefficient $T_{h e }$ appears, with peaks emerging around the energy region corresponding to the bottom or the top of subbands in Fig. \ref{Fig2}(a). Furthermore, the inner peaks also reflect feature of SC gap controlled by altermagnet.
The spin-flip reflection coefficient $T_{{\bar e}e}$ exhibits characteristics similar to $T_{he}$, as does the normal Andreev reflection coefficient $T_{{\bar h}e}$. However, the latter contributes nothing to spin transport because the incident electron is reflected as an opposite-spin hole,
equivalent to injecting a spin-singlet Cooper pair into SC.
{The normal reflection coefficient $T_{{e}e}$ is absent around the energy region corresponding to the bottom or the top of subbands in Fig. \ref{Fig2}(a).}
The total reflection coefficient equals unity within the SC gap and decreases outside the gap due to the quasiparticle tunneling.
Figures \ref{Fig2}(c) and \ref{Fig2}(d) illustrate parameter dependence of $T_{he}$.
As illustrated in Fig. \ref{Fig2}(c), $T_{he}$ can be significantly enhanced by tuning the chemical potential $\mu$, indicating a strong spin-triplet component in the superconducting nanowire.
Besides, increasing $\mu$ could lead to enhanced peak separation.
The underlying mechanism can be explained as follows.
As illustrated in Fig. \ref{Fig1}(a), when increasing the chemical potential, altermagnetic energy band is lowered, exhibiting more pronounced spin splitting near the Fermi level.
Therefore, the SC-proximity-induced energy band in Fig. \ref{Fig2}(a) would exhibit a stronger spin splitting and a reduced SC gap, resulting in broader peak intervals in the $T_{he}$ spectrum of positive-energy or negative-energy ranges.
Figure \ref{Fig2}(d) exhibits a similar dependence on altermagnet strength, following analogous physical mechanisms. The increased $t_J$ induces more pronounced band splitting, resulting in widened peak spacings.
We then select four points in Fig. \ref{Fig2}(c) [blue cross points] to study the polar angle dependence of equal-spin Andreev reflection coefficient.
As illustrated in Fig. \ref{Fig3}, the $T_{he}$ reaches its maximum when the incident electron spin is aligned along the direction perpendicular to $z$, and diminishes to zero as the spin orientation rotates toward the $\pm z$ direction, consistent with our previous analysis in Eq. (\ref{E11}).

Next, we investigate the spin transport characteristics. We select the spin bias along $x$ direction as a representative case.
In Fig. \ref{Fig4}(a), we study the dependence of spin conductance on the spin bias for different chemical potentials.
The peak spacings expand and the peak amplitudes increase when increasing chemical potential.
As derived from Eq. (\ref{E7}), spin conductance is contributed by $T_{{\bar e}e}$, $T_{he}$ and $T_{trans}$.
Recall that Fig. \ref{Fig2}(b) reveals that $T_{{\bar e}e}$ and $T_{he}$ have similar dependence on incident electron energy and $T_{trans}$ remains finite outside the SC gap.
Based on these findings, we can conclude that the evolution of spin conductance peaks with chemical potential depends on both $T_{{\bar e}e}$ and $T_{he}$.
When spin bias is smaller than the SC gap, spin conductance is contributed only by $T_{{\bar e}e}$ and $T_{he}$ \cite{mao_Charge_2021a}.
When the spin bias is larger than the SC gap, quasiparticle tunneling also contributes to spin conductance, with its contribution growing when the chemical potential increases.

We focus on the spin transport with spin bias inside the SC gap.
We select the blue cross point in $\mu=80\Delta$ curve in Fig. \ref{Fig4}(a) and rotate the direction of spin bias to study the polar angle dependence of spin conductance. As illustrated in Fig. \ref{Fig4}(b), the spin conductance reaches its maximum when the spin bias is oriented perpendicular to the $\pm z$ direction. $G_{s,tot}$ diminishes to zero when spin bias rotates toward the $\pm z$ direction due to the absence of both equal-spin Andreev reflection and spin-flip reflection processes. When spin bias aligns with Néel vector of the altermagnet, the spin quantization axis coincides with the $z$-direction. Consequently, the spin splitting term in spin space becomes diagonal and spin $z$ is maintained as a good quantum number, which effectively forbids the spin-flip processes at the interface.
Note that the spin current contributed by spin-flip reflection is usually dissipated at the interface rather than being injected into nanowire \cite{ dai_Spin_2022a}, so we calculate the spin injection efficiency using Eq. (\ref{E8}) and investigate the polar angle dependence of spin injection efficiency in Fig. \ref{Fig4}(c).
According to Eq. (\ref{E8}), spin injection efficiency in Fig. \ref{Fig4}(c) times total spin conductance in Fig. \ref{Fig4}(b) represents the spin conductance corresponding to the supercurrent injected into the superconducting nanowire.
As illustrated in Fig. \ref{Fig4}(c), the efficiency is almost angle-independent. This is because the angle-dependence of equal-spin Andreev reflection and spin-flip reflection is similar.
As a result, the dissipated part $G_{s,dis}$ of the differential spin conductance depends on the polar angle in almost the same way as the total spin conductance $G_{s,tot}$, then the spin injection efficiency is almost angle-independent.
In addition, due to the $\sigma_z$ symmetry of the Hamiltonian $H^{BdG}$ in Eq.(\ref{E1}) at the absence of SOC, the spin conductance and the spin injection efficiency are independent of the azimuthal angle $\phi$, as demonstrated in Figs. \ref{Fig4}(b, c).

\section{\label{sec4}Spin transport in the presence of SOC}

\begin{figure}[!htb]
\centerline{\includegraphics[width=\columnwidth]{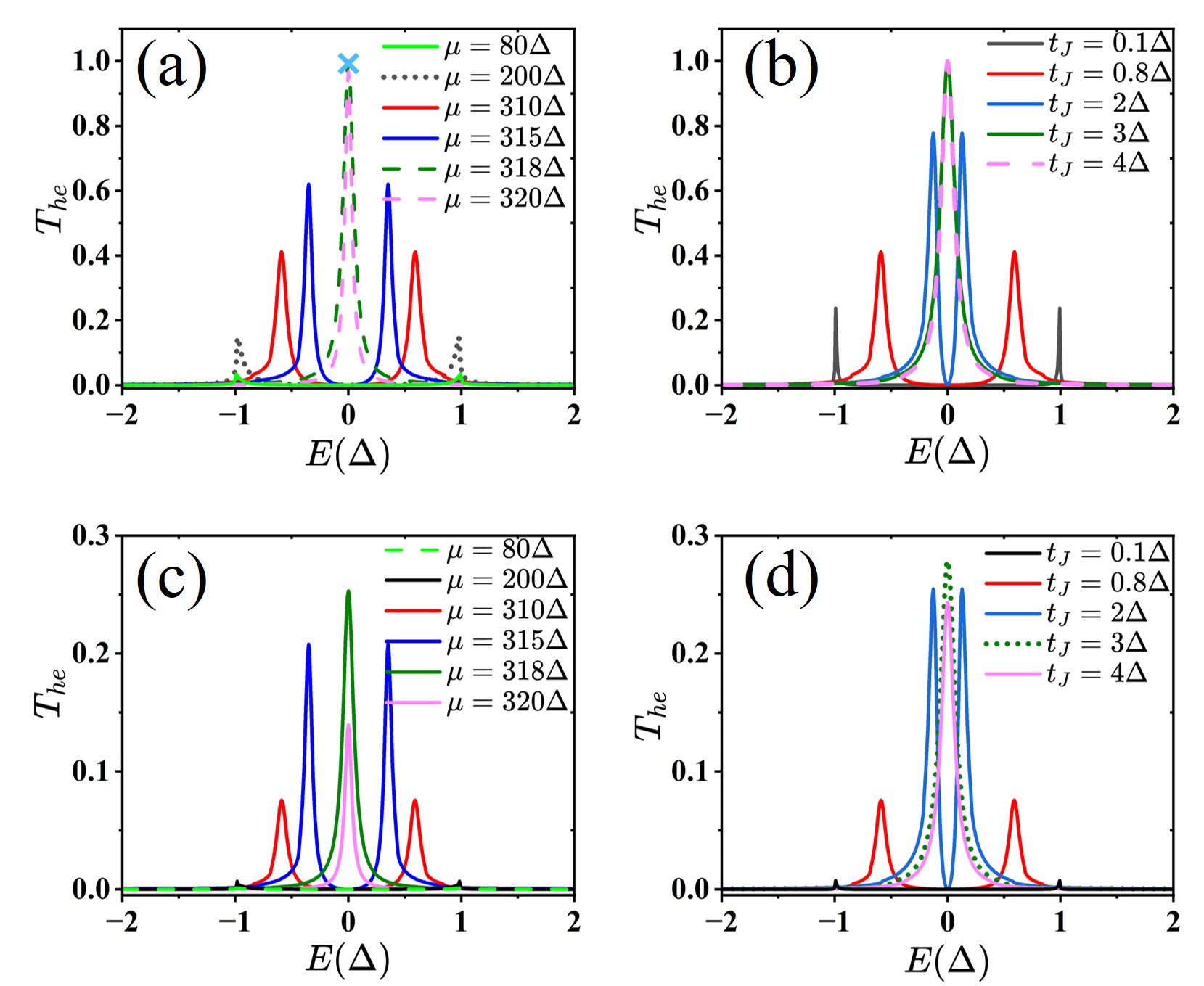}}
\caption{
    Equal-spin Andreev reflection coefficient $T_{he}$ versus incident electron energy $E$ for (a,c) different chemical potentials $\mu$ and (b,d) different altermagnet strengths $t_J$. The incident electron spin is along $z$ axis in (a-b) and $x$ axis in (c-d). Parameters: $\alpha=11.2\Delta$ and $\phi=0$ in (a-d), $t_J=0.8\Delta$ in (a,c), $\mu=310\Delta$ in (b,d), $\theta=\pi/2$ in (c-d) and $\theta=0$ in (a-b).
}\label{Fig5}
\end{figure}

In this section, we thoroughly study the spin transport characteristics in the presence of SOC.
Through Eq. (\ref{E10}), we obtain $\bm{d}=(d_x,d_y,d_z)$, where three components are all nonzero, indicating the spin-triplet pairing exists in 1D nanowire with $(\ket{\uparrow\downarrow}+\ket{\downarrow\uparrow})_{x,y,z}$ \cite{tang_Magnetic_2021a}.
This changes the constraint on spin orientation for the existence of equal-spin pairing Cooper pairs.
We then investigate the %parameters dependence of
equal-spin Andreev reflection coefficient and select incident electron spin aligned with $z$ direction as a representative case.
As illustrated in Figs. \ref{Fig5}(a) and \ref{Fig5}(b), the finite $T_{he}$ displays the nonzero equal-spin pairing component.
In Sec. \ref{sec3}, where spin-orbit coupling (SOC) was absent, the $\bm{d}$ vector has only a $d_z$ component, thus electrons with spin $z$ could not undergo equal-spin Andreev reflection.
But here, with the introduction of SOC, the $\bm{d}$ vector acquires $d_x$ and $d_y$ components.
This enables the formation of spin $zz$ Cooper pairs, allowing $z$-spin electrons to participate in equal-spin Andreev reflection and thereby inject a $z$-direction spin-polarized supercurrent.
Figure \ref{Fig5}(a) presents the $T_{he}$ for different chemical potentials.
The peaks correspond to the bottom or the top of the subbands.
Notably, unlike the case shown in Fig. \ref{Fig2}(c),
the SC gap here is protected by SOC \cite{oreg_Helical_2010a}, showing weaker reduction under the same parameter conditions with Fig. \ref{Fig2}(c).
With further increase of $\mu$, the system experiences a topological phase transition and enters into topological superconducting phase, which requires $\mu$ and $t_J$ to satisfy $(4t_J)^2> \Delta^2+(4t-\mu)^2$. The details of topological phase transition are presented in Appendix \ref{appendixA}. In Fig. \ref{Fig5}(a), the critical value for the transition is approximately $\mu_m=317\Delta$. Before the transition, the gap gradually closes, accompanied by approach of $T_{he}$ peaks.
When $\mu>\mu_m$, although the SC gap remains open, the $T_{he}$ peak is pinned at zero energy with its amplitude close to 1 because of resonant Andreev reflection from Majorana zero mode \cite{he_Selective_2014a,law_Majorana_2009a}.
This phenomenon can also be understood as follows: due to the spin splitting of altermagnet, when $\mu>\mu_m$, almost only spin-up electrons exist near the Fermi level. Thus, the equal-spin Cooper pairs $\ket{\uparrow \uparrow}_z$ are dominant in the superconducting nanowire.
Consequently, the equal-spin Andreev reflection becomes the dominant scattering process with the spin-flip reflection coefficient remains finite but significantly smaller than $T_{he}$.
Figure \ref{Fig5}(b) presents the $T_{he}$ for different altermagnet strengths. The critical value for the phase transition is approximately $t_{Jm}=2.5\Delta$. Similar to Fig. \ref{Fig5}(a), before the transition, $T_{he}$ peaks gradually approach. When $t_J>t_{Jm}$, the $T_{he}$ peak is pinned at zero energy with its amplitude close to 1.

\begin{figure}[!htb]
\centerline{\includegraphics[width=\columnwidth]{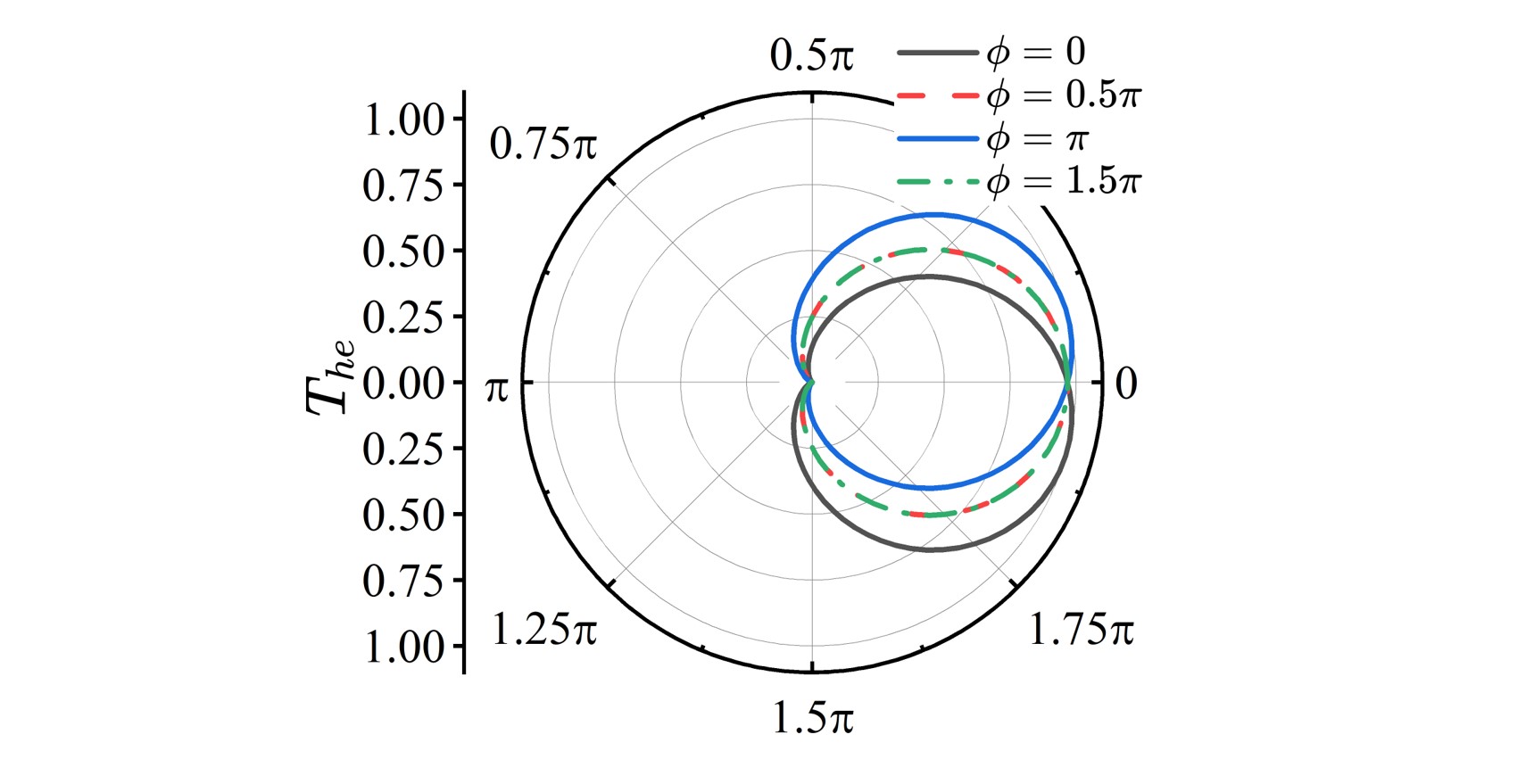}}
\caption{
    Equal-spin Andreev reflection coefficient $T_{he}$ versus polar angle $\theta$ for different azimuthal angles $\phi$. The point at $\theta=0$ on $\phi=0$ curve corresponds to blue cross point in Fig. \ref{Fig5}(a). Parameters: $\alpha=11.2\Delta$, $t_J=0.8\Delta$, $\mu=320\Delta$ and $E=0$.
}\label{Fig6}
\end{figure}

Under the same conditions, we calculate equal-spin Andreev reflection coefficient for the incident electron with spin along $x$ direction.
As illustrated in Figs. \ref{Fig5}(c) and \ref{Fig5}(d), the results exhibit similar characteristics to those presented in Figs. \ref{Fig5}(a) and \ref{Fig5}(b).
The magnitudes of peaks differ because the corresponding $xx$- and $zz$-spin Cooper pairs match differently with the Cooper pairs in superconducting nanowire.
We select the blue cross point in Fig. \ref{Fig5}(a) to systematically study the direction dependence of equal-spin Andreev reflection coefficient.
As illustrated in Fig. \ref{Fig6}, the $T_{he}$ reaches its maximum when the incident electron spin is oriented close to the $+z$ direction, and diminishes to zero as the spin orientation
rotates toward near $-z$ direction.
This is because the Cooper pairs are almost $\ket{\uparrow \uparrow}_z$ pairs in the topological superconducting phase, prohibiting the injection of Cooper pairs with $\ket{\downarrow \downarrow}_z$ spin.
In addition, under the presence of SOC, the $\sigma_z$ symmetry is broken, which causes the equal-spin Andreev reflection coefficient to depend on the azimuthal angle $\phi$ (see Fig. \ref{Fig6}).

\begin{figure}
\centerline{\includegraphics[width=\columnwidth]{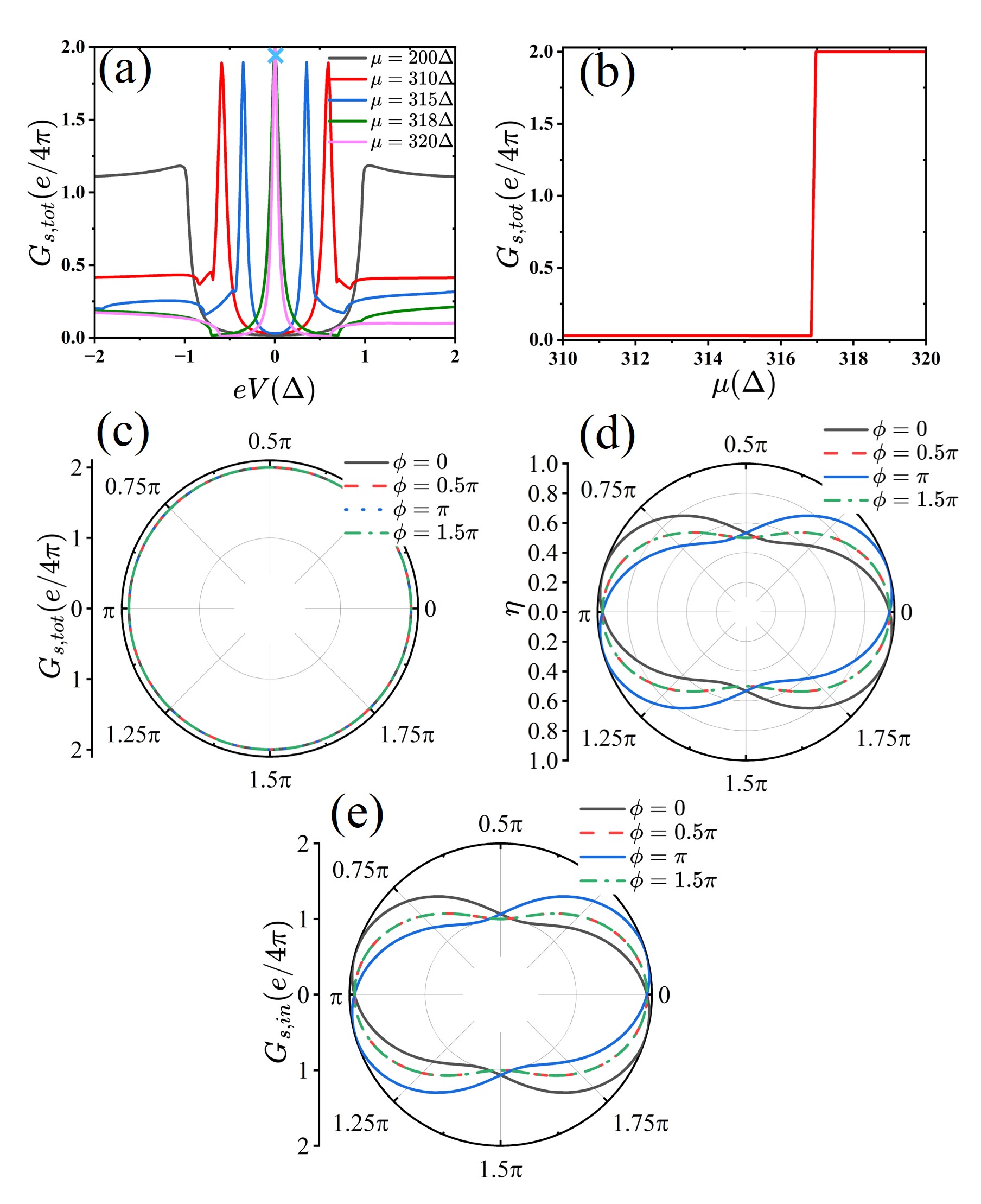}}
\caption{
    (a) Spin conductance $G_{s,tot}$ versus spin bias $V$ for different chemical potentials $\mu$. (b) Spin conductance $G_{s,tot}$ versus chemical potential $\mu$. Point at $\mu=320\Delta$ corresponds to blue cross point in (a). (c) Total spin conductance $G_{s,tot}$ versus polar angle $\theta$ for different azimuthal angles $\phi$. Point at $\theta=0$ on $\phi=0$ curve corresponds to blue cross point in (a). (d) Spin injection efficiency versus polar angle $\theta$ for different azimuthal angles $\phi$. (e) Injected spin conductance $G_{s,in}$ versus polar angle $\theta$ for different azimuthal angles $\phi$. Parameters: $\alpha=11.2\Delta$ and $t_J=0.8\Delta$ in (a-e), $\mu=320\Delta$ in (c-e), $\theta=0$ and $\phi=0$ in (a-b), $V=0$ in (b-e).
}\label{Fig7}
\end{figure}

Next, we investigate the spin transport in the presence of SOC and select the spin bias along $z$ direction as a representative case. In Fig. \ref{Fig7}(a), we study the spin conductance as
a function of the spin bias for different chemical potentials.
The $\mu=200\Delta$ curve demonstrates that the conductance gap remains pinned at approximately $eV=\pm\Delta$ with $\mu$ being moderate and far from transition point. However, peaks are not obvious in this gap region because the equal-spin Andreev reflection and spin-flip reflection are weak.
When $\mu$ continues to increase, the contribution of $T_{{\bar e}e}$ and $T_{he}$ increases, leading to the emergence of remarkable peaks.
During phase transition, the peaks gradually approach and eventually become pinned at zero spin bias when $\mu>\mu_m$, with its magnitude reaching a quantized value of $e/2\pi$.
Then we select the blue cross point at zero spin bias in $\mu=320\Delta$ curve in Fig. \ref{Fig7}(a) and study other parameter dependence in Figs. \ref{Fig7}(b) and \ref{Fig7}(c).
Here because we study $V=0$ case, the spin conductance is only contributed by $T_{{\bar e}e}$ and $T_{he}$.
Figure \ref{Fig7}(b) reveals a pronounced shift in zero-bias spin conductance at the topological phase transition point. Before the transition, it is nearly zero, but after the transition, it reaches a quantized value of $e/2\pi$, which is consistent with our analysis above.
Figure \ref{Fig7}(c) illustrates the direction dependence of spin conductance, which maintains a quantized value of $e/2\pi$ at all directions of spin bias.
The spin conductance originates from the equal-spin Andreev reflection and spin-flip reflection processes with the spin orientations ${\hat{\bm{n}}}$ and $-{\hat{\bm{n}}}$. Although these coefficients are dependent on the spin direction of incident electrons, their total contributions to spin conductance are independent of the direction of spin bias ${\hat{\bm{n}}}$.
This result is consistent with the literature \cite{mao_Charge_2021a}, arising from the topological superconducting phase and Majorana zero mode.
However, due to the $T_{{\bar e}e}$'s sensitivity to spin orientation, the spin injection efficiency exhibits strong dependence on the direction of the spin bias.
As shown in Fig. \ref{Fig7}(d), when the spin bias lies within the $xoy$ plane, $T_{{\bar e}e}$ is large, leading to a diminished spin injection efficiency at $\theta=0.5\pi$. In contrast, when the spin bias aligns along $\pm z$ direction, spin conductance is mainly contributed by the nearly quantized $T_{he}$, leading to a significantly enhanced spin injection efficiency near $\theta=0$ and $\theta=\pi$, with the maximum spin injection efficiency reaching 1.
In Fig. \ref{Fig7}(e), we also calculate the injected spin conductance $G_{s,in}=G_{s,tot}-G_{s,dis}$ according to Eqs. (\ref{E7}) and (\ref{E8}), also equalling the spin injection efficiency in Fig. \ref{Fig7}(d) times total spin conductance in Fig. \ref{Fig7}(c). When the spin bias aligns near $\pm z$ direction, the injected spin conductance contributed by the equal-spin Andreev reflection reaches the nearly quantized value of $e/2\pi$, also indicating the suppression of other reflection processes. When the spin bias lies in the $xoy$ plane, the injected spin conductance is smaller.

\section{\label{sec5}Discussion and Conclusion}
It is well known that the superconductor-ferromagnet structure can host spin-triplet superconductivity \cite{bergeret_Odd_2005}, whereas our work focuses on the superconductor-altermagnetic nanowire structure. It is thus instructive to briefly discuss their similarities and differences. Regarding the similarities, if we replace the altermagnetic term ($T_j k_x^2 \sigma_z$) in the effective Hamiltonian in Eq. (\ref{E1}) with a ferromagnetic one ($B_z \sigma_z$), which corresponds to a ferromagnetic wire with proximity-induced superconductivity, we can still obtain a nonzero ${\bm{d}}$ vector according to Eq. (\ref{E10}). The nonzero ${\bm{d}}$ vector signifies the spin-triplet pairing. In summary, in both systems, the necessary ingredient to induce spin-triplet pairing is the spin splitting term, the specific form of which (altermagnetic or ferromagnetic) does not affect the emergence of spin-triplet pairing.
About the differences,
magnetism in ferromagnets is generally detrimental to superconductivity. However, altermagnets exhibit zero net magnetism while still possessing spin splitting, and are beneficial for preserving the superconductivity. In particular, the spin-splitting of altermagnet, described by the $d$-wave like term $k_x^2-k_y^2$, is not only momentum magnitude-dependent but also direction-dependent. As a result, the hybrid system composed by altermagnet and $s$-wave superconductivity can exhibit anisotropic spin-triplet strength \cite{wei_gapless_2024, maeda_Classification_2025}. Thus, when a one-dimensional nanowire is proximitized to such a system, the magnitudes of the induced altermagnetism ($T_j$) and resulting spin-triplet component vary with the orientation of the placed nanowire, and can even vanish for certain directions. This renders the spatial direction an extra degree of freedom for control. 

In conclusion, we demonstrate that spin-triplet superconductivity and spin supercurrent can be generated in a nanowire in proximity to altermagnet and $s$-wave SC.
Using the pairing correlation function, we obtain equal-spin pairing correlation in both cases, whether SOC is absent or present.
Using the nonequilibrium Green's function method, we calculate equal-spin Andreev reflection coefficient and verify the injection of spin-triplet Cooper pairs.
The zero resistance and dissipationless nature of Cooper pairs support the long-range propagation of spin supercurrent.
We also systematically investigate the spin conductance and spin injection efficiency.
We show that spin transport performance can be significantly regulated by a topological phase transition.
These results are beneficial to the further development of superconducting spintronics
and the design of spin devices.

\begin{acknowledgments}
We thank Peng-Yi Liu for fruitful discussions.
This work was financially supported by the National Key R and D Program of China (Grant No. 2024YFA1409002), the National Natural Science Foundation of China (Grant No. 12374034), 
the Quantum Science and Technology-National Science and Technology Major Project (Grant No. 2021ZD0302403), and the China National Postdoctoral Program for Innovative Talents (Grant No. BX20250182). The computational resources are supported by High-performance Computing Platform of Peking University.
\end{acknowledgments}

\appendix
\section{\label{appendixA} Topological phase transition}

\begin{figure}[h]
\centerline{\includegraphics[width=\columnwidth]{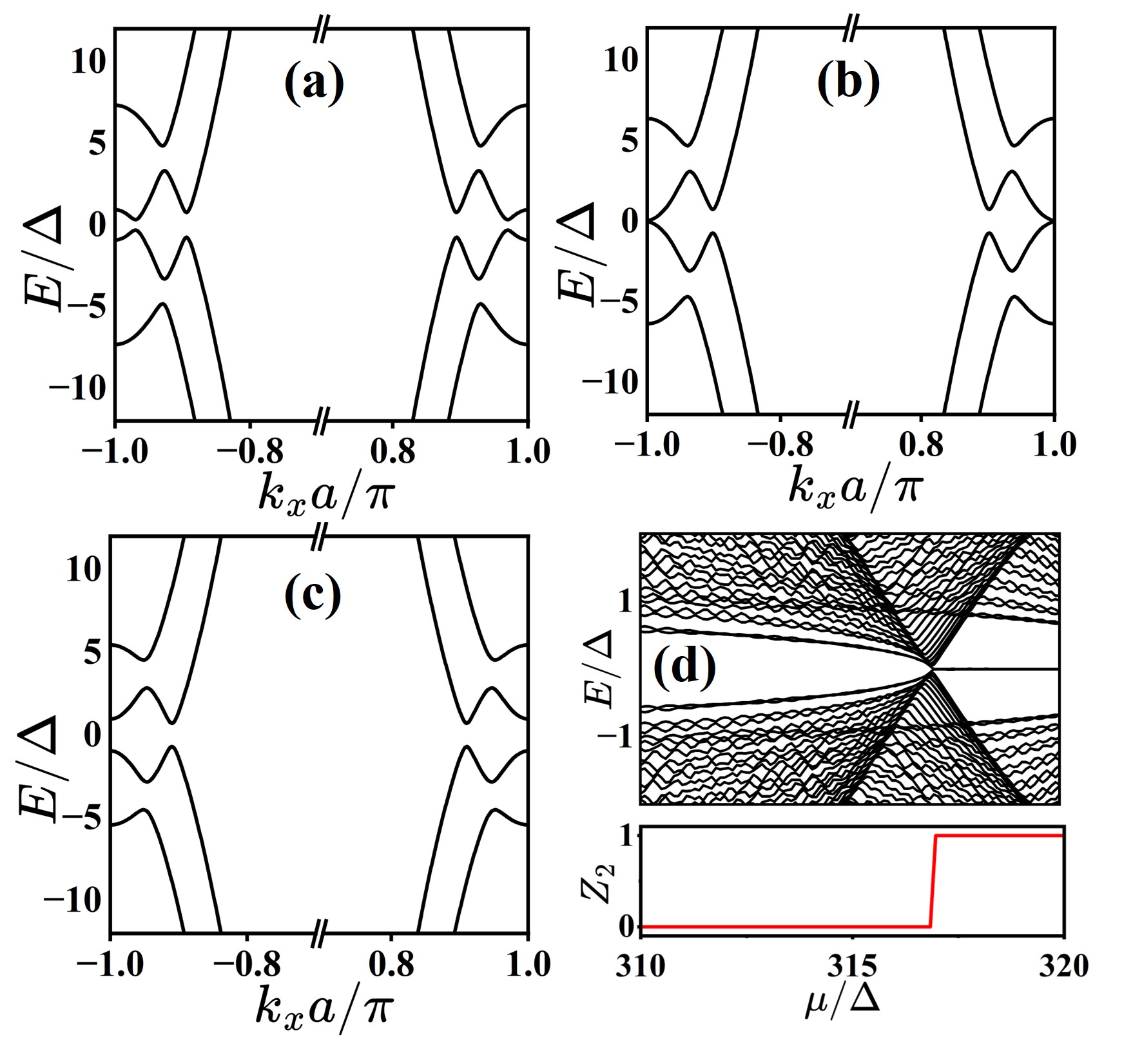}}
\caption{
    (a-c) Band inversion process 
   with the chemical potential $\mu=316\Delta$ in (a), 
   $317\Delta$ in (b) and $318 \Delta$ in (c).
    (d) Top panel: Finite-size energy spectrum of altermagnetic superconducting nanowire versus $\mu$. Bottom panel: $Z_2$ topological invariants of the altermagnetic superconducting nanowire versus $\mu$. Parameters: $\alpha=11.2\Delta$ and $t_J=0.8\Delta$ in (a-d).
}\label{Fig8}
\end{figure}

The Hamiltonian of altermagnetic superconducting nanowire in Eq. (\ref{E3}) can be Fourier transformed as
\begin{align}
    H_{nw}&=\sum_{k_x,s} \left[ 2t-2t\cos(k_x a)-\mu\right] c_{k_x,s}^\dagger c_{k_x,s}  \nonumber \\  
    &+\sum_{k_x,s,s'} \left[2t_J-2t_J\cos(k_x a)  \right] c_{k_x,s}^\dagger(\sigma_z)_{s,s'} c_{k_x,s'}  \nonumber \\  
    &+\sum_{k_x,s,s'} \alpha \sin(k_x a) c_{k_x,s}^\dagger(\sigma_y)_{s,s'} c_{k_x,s'} \nonumber \\  
    &+\sum_{k_x} (\Delta c_{k_x,\uparrow}^\dagger c_{-k_x,\downarrow}^\dagger +H.c. )
    ,
    \label{A1}
\end{align}
where $c_{k_x,s}^\dagger(c_{k_x,s})$ is the creation (annihilation) operator of electron with spin $s$ and momentum $k_x$.
Using Eq. (\ref{A1}), we calculate the energy spectrum and present the phase transition process in Figs. \ref{Fig8}(a-c), manifested by the closure and subsequent reopening of the energy gap at the high-symmetry point $k_x a=\pi$ when the chemical potential $\mu$ increases. The gap closes at approximately $\mu=317\Delta$.
To derive the gap-closing criterion, we rewrite the Hamiltonian in Eq. (\ref{A1}) within Nambu basis as matrix %$H_{nw}=\sum_{k_x} H_{nw}(k_x)$ and
\begin{align}
    H_{nw}(k_x)=
	 \begin{bmatrix}
  T_2+J_2 &-iA_2&0&\Delta  \\
  iA_2&T_2-J_2&-\Delta&0  \\
  0&-\Delta&-T_2-J_2&iA_2 \\
  \Delta &0&-iA_2&-T_2+J_2
\end{bmatrix}
	,       \label{A2}
\end{align}
where we have set $T_2=2t-2t\cos(k_x a)-\mu$, $J_2=2t_J-2t_J\cos(k_x a)$ and $A_2=\alpha \sin(k_x a)$. By diagonalizing the Hamiltonian matrix, the energy eigenvalues are obtained 
 \begin{align}
	E_{\pm}^2=T_2^2+A_2^2+J_2^2+\Delta^2\pm 2\sqrt{\Delta^2 J_2^2+T_2^2(A_2^2+J_2^2)},
	      \label{A3}
\end{align}
where $E_{+}$ denotes two bands far from zero energy and $E_{-}$ denotes two bands near zero energy. At $k_x a=\pi$, $A_2$ vanishes. Utilizing the gap-closing condition ($E_{-}(k_x a=\pi)=0$), we obtain $J_2^2=T_2^2+\Delta^2$. By substituting $T_2=2t-2t\cos(k_x a)-\mu$ and $J_2=2t_J-2t_J\cos(k_x a)$, we obtain the phase transition criterion $(4t_J)^2= (4t-\mu)^2+\Delta^2$. This explains our gap closing in Figs. \ref{Fig8}(a-c).

Using the Hamiltonian in Eq. (\ref{E3}), we obtain the eigenenergy of finite-size altermagnetic superconducting nanowire versus the chemical potential $\mu$, as demonstrated in the top panel of Fig. \ref{Fig8}(d). %the top panel of Fig. \ref{Fig8}(d) demonstrates the eigenenergy of finite-size 1D altermagnetic superconducting nanowire versus the chemical potential $\mu$. 
After the gap closing and reopening at approximately $\mu=317\Delta$, the Majorana zero modes with zero energy emerge. Furthermore, using the Pfaffian method \cite{fu_Time_2006} and the Hamiltonian [Eq. (\ref{A1})], we also calculate $Z_2$ topological invariants versus $\mu$ in the bottom panel. 
The $Z_2=1$ region confirms the result in the top panel, validating the existence of the Majorana zero modes. The trend of $Z_2$ curve is consistent with result of the total spin conductance in Fig. \ref{Fig7}(b).

\bibliography{refer.bib}
\end{document}